\documentclass[]{aa}
\usepackage{graphicx}
\usepackage{subfigure}
\usepackage{amsmath}
\usepackage{amssymb}
\usepackage{txfonts}
\def\kms{\relax \ifmmode {\,\mbox{km\,s}}^{-1}\else \,\mbox{km\,s}$^{-1}$\fi}
\def\ha{\relax \ifmmode {\mbox H}\alpha\else H$\alpha$\fi}
\def\hb{\relax \ifmmode {\mbox H}\beta\else H$\beta$\fi}
\def\hi{\relax \ifmmode {\mbox H\,{\scshape i}}\else H\,{\scshape i}\fi}
\def\hii{\relax \ifmmode {\mbox H\,{\scshape ii}}\else H\,{\scshape ii}\fi}
\def\heii{\relax \ifmmode {\mbox He\,{\scshape ii}}\else He\,{\scshape ii}\fi}
\def\oiii{\relax \ifmmode {\mbox O\,{\scshape iii}}\else O\,{\scshape iii}\fi}
\def\oii{\relax \ifmmode {\mbox O\,{\scshape ii}}\else O\,{\scshape ii}\fi}
\def\oi{\relax \ifmmode {\mbox O\,{\scshape i}}\else O\,{\scshape i}\fi}
\def\nii{\relax \ifmmode {\mbox N\,{\scshape ii}}\else N\,{\scshape ii}\fi}
\def\sii{\relax \ifmmode {\mbox S\,{\scshape ii}}\else S\,{\scshape ii}\fi}
\def\siii{\relax \ifmmode {\mbox S\,{\scshape iii}}\else S\,{\scshape iii}\fi}
\def\lha{\relax \ifmmode \mbox {L}_{H\alpha}\else $\mbox{L}_{H\alpha}$\fi}
\def\ldig{\relax \ifmmode {\mbox L}_{DIG}\else ${\mbox L}_{DIG}$\fi}
\def\ls{\relax \ifmmode {\mbox L}_{ Str}\else ${\mbox L}_{ Str}$\fi}
\def\eme{\relax \ifmmode {\,\mbox{pc\,cm}}^{-6}\else \,pc\,cm$^{-6}$\fi}
\def\l{\relax \ifmmode  \lambda\else $\lambda$\fi}

\def\arcmin{\hbox{$^\prime$}}
\def\arcsec{\hbox{$^{\prime\prime}$}}
\def\deg{\hbox{$^\circ$}}
\def\fs{\hbox{$^{\rm s}$}}
\def\hms#1h#2m#3s{\relax \ifmmode #1^{\rm h}\,#2^{\rm m}\,#3^{\rm s}
                   \else \hbox{$#1^{\rm h}\,#2^{\rm m}\,#3^{\rm s}$}
                  \fi}
\def\dms#1d#2m#3s{\relax#1\deg\,#2\arcmin\,#3\arcsec}
\def\hmsd#1h#2m#3.#4s{\relax\ifmmode #1^{\rm h}\,#2^{\rm m}\,#3.#4\fs
                      \else \hbox{$#1^{\rm h}\,#2^{\rm m}\,#3#4\fs$}
                      \fi}
\begin{document}

   \title{Energy balance in two phase models for temperature fluctuations in {\hii} regions.}


   \author{C. Giammanco \inst{1}
          \and J.E. Beckman \inst{1,2}
	   }

   \offprints{C. Giammanco}

   \institute{Instituto de Astrof\'\i sica de Canarias, C. V\'\i a L\'actea s/n,
      38200, La Laguna, Tenerife, Spain\\
              \email{corrado@ll.iac.es}
         \and
             Consejo Superior de Investigaciones Cient\'{\i}ficas, Spain
             }

   \date{}

\abstract{{\it Aims}. The main objective of this article is to provide a simple physical framework with permits a quantitative comparison of measurements of the temperature fluctuations in the ionized interstellar medium with possible mechanisms which can produce them. \\ 
{\it Methods}. We assume a generalized two phase ISM and derive expressions  relating the mean amplitude of the temperature fluctuations to the temperatures of the phases and to the energy input, in excess of the basic component due to photoionization, required to maintain them. We apply these expressions to a set of limiting cases for the temperature and density differences between the phases. Finally we compare the most plausible case with the most complete data set available: the temperature fluctuations observed in the Orion Nebula.\\
{\it Results}. We first list the cases considered and our general inferences:
(a) Very hot tenuous substrate and warm moderately dense clouds; discarded
as requiring too much excess energy input.
(b) Two phases with equal densities but different temperatures, both warm; feasible but not general.
(c) Two warm phases at moderately different temperatures and densities; 
the most probable case. This case is then used to quantify a specific 
hypothesis, reconnection of turbulent magnetic fields, as the source of the
fluctuations observed in the Orion Nebula. Field strengths of a few 
hundred $\mu$G are required, not out of line with the limited observations 
available. Time variability on scales of months is a testable prediction 
of the scenario.  
     }
 \maketitle

\section{Introduction}
The presence of temperature fluctuations is an interesting problem which remains to be resolved in the field of physics of the ionized interstellar medium. Temperature fluctuations arise in planetary nebulae, and in {\hii} regions from the smallest to the most luminous. Although it is not surprising to
find inhomogeneities either in density or in temperature within a medium such
as the ISM, we would like to be in a position to account physically for these
inhomogeneities, especially as conventional photoionization models  yield
temperature fluctuations considerably smaller than those observed (Stasi\'nska, 2000).
The  objects mentioned above are very varied in their sizes, chemical
composition, morphology, and not least in their origins, so that it might well
be the case that different mechanisms operate within them to produce the
temperature fluctuations observed, and in the literature we can find a number
of different suggestions for these. Torres--Peimbert et al. (1990) proposed
that local variations in chemical composition might give rise to the
temperature fluctuations seen in planetary nebulae. For a specific ``giant"
extragalactic {\hii} region (NGC 2363) Luridiana et al. (2001) explored the
possibility that the stellar winds from the ionizing stars might give rise
to the temperature fluctuations observed, but in the end rejected this
hypothesis as the mechanism was incapable of supplying sufficient energy. The effects of stellar winds have been previously studied by Peimbert et al. (1991), and Esteban (2002) examine all these possible causes of temperature fluctuations in a comprensive review.\\
   Before looking in any detail at specific mechanisms it is worth
drawing a distinction between smaller {\hii} regions, such as the Orion nebula, ionized by one or only a few stars, and the luminous regions, often termed
``giant", regions ionized by one or even more major star clusters. The Orion
nebula is an ionized bubble or blister at the edge of a much more massive
molecular cloud complex. If its ionizing stars were much more luminous, they
would ionize a much greater proportion of this complex, which we can resolve
into numerous cloud components thanks to its proximity. Under these
circumstances, if viewed from an external galaxy, this hypothetical zone would
look like a single giant {\hii} region, but we know that it would be composed
of a mixed group of photoionized and neutral zones (see e.g. Giammanco et al., 2004, 2005). Within this structure it is quite plausible that the neutral zones
which are too dense for penetration by substantial amounts of UV radiation
will be weakly ionized by cosmic rays. This proposal was quantified in
Giammanco \& Beckman (2005, Paper I) where we used it to derive the values of the
temperature fluctuations measured in this type of regions. However even if
this hypothesis is valid, it can apply only to these large multi--phase regions,
since it depends on the density and temperature contrast between the cool
weakly ionized dense clumps and the warm fully ionized interclump medium.
In the Orion nebula we observe only the photoionized medium, so the temperature
fluctuations must have a different origin. Even so the formalism proposed
in Paper I in our study of the giant regions is
independent of the physical mechanism underlying the fluctuations, and can be
used as the basis for a more general study.\\
In the present article we base our work on this formalism to explore a number of possibilities for the production of the observed amplitudes of the temperature fluctuation parameter
$t^2$, as defined by Peimbert (1967). The study is based essentially
on energy considerations, and in this sense has some similarities to work
published by Binette et al.~(2001), and Luridiana et al. (2001). However those authors used a statistically defined temperature distribution as the basis for their models, while we
will look at conceptually simpler two phase models, which give the advantage
that they are easier to formulate analytically, thus permitting us to obtain general
equations which can be used to define and examine extreme cases. These include
the limiting cases of large and small temperature fluctuations, and also of large
and small masses of gas involved in the temperature excursions. It is also
straightforward in this formulation to obtain reasonable approximate estimates
of how much energy will be needed for a given physical process to produce an
observed value of $t^2$. In the first part of the article we will derive and
describe explicitly the equations to be used in the model. We will then go on
to apply these to a set of limiting cases which illustrate the effects of
varying the physical parameters, and will be used to eliminate a number of
possible mechanisms as well as set limiting conditions on others. Finally
we will present our conclusions and a suggestion for testing a specific
possible mechanism, that of magnetic field reconnection, as the source of
the observed fluctuations as applied to the case of the Orion nebula.\\   
In a previous article Paper I
we showed how a model of the interstellar medium (ISM) comprising two phases
with a strong density contrast, some two orders of magnitude, could be
used as the basis for an explanation of the spatial fluctuations in
temperature measured in {\hii} regions via observed emission line ratios.
The model required a specific mechanism to supply the local ionization which
gives rise to the line emission from the denser cooler phase, and we invoked
the presence of cosmic ray protons accelerated in the dynamic environment of
the hot stars within the regions as a suggestion for this mechanism. In the
present article we generalize the model so that it can accommodate a variety of density and temperature regimes, and also a variety of possible mechanisms
for the ``extra" local ionization, over and above the general photoionization
which is characteristic of all {\hii} regions. We then consider a specific
mechanism by way of an example, that of magnetic field line reconnection, and
show that it is quantitatively compatible with the most complete set of published observations in the context
of our models.

\section{The equations of the generalized model.}
\subsection{An initial version: some simple cases.}
\label{s0}
In Paper I we developed the basic equations for the two phase
model which we quote here to initiate our development:
\begin{equation}
T_{\rm c}= T_{\circ}\;  \left(1-\sqrt{\frac{{t^2}}{x^2\theta}}\right),
\label{p1}
\end{equation}
\begin{equation}
T_{\rm w}=T_{\circ}\;  \left(1+\sqrt{t^2 x^2\theta}\right).
\label{p2}
\end{equation}
where $T_{\rm c}$ and  $T_{\rm w}$ are respectively the temperatures of the cool and the warm phase, $T_{\circ}$ is the mean temperature, $\theta$ is the ratio of the masses in the cool and warm phases, $x$ is the ionization fraction of the ions in the cool phase and $t^2$ is the measured temperature fluctuations. $t^2$ and $T_{\circ}$ were introduced by Peimbert (1967) and we defined then explicitly in Paper I.\\
In the present article we assume that the emitting material
is virtually entirely photoionized in both phases so that $x\sim 1$.
We also assume that in the absence of an extra heating mechanism the
whole of the medium would reach a temperature $T_{\rm c}$, in equilibrium 
between the radiative heating accompanying photoionization and 
the radiative cooling mechanisms which operate. The temperature 
fluctuations are caused by a mechanism (not specified at this point)
which can raise the temperature of a fraction of the total mass.
Our first aim is to quantify the energy required of this process
in order to generate a given value of $t^2$. If the temperature of the gas in the absence of the additional process is $T_{\rm c}$, an additional energy $\Delta E$ is required to push the temperature of a fraction of the gas up to $T_{\rm w}$:

\begin{equation}
\Delta E= \frac{3}{2}\,K\,  N_{\rm w}\, T_{\circ}\;  \left(\sqrt{t^2\theta}+\sqrt{t^2/\theta}\right).
\label{p3}
\end{equation}

where $N_{\rm w}$ is the total number of electrons in the phase with           
temperature $T_{\rm w}$. If $N$ is the total number of electrons in the medium
we can rewrite  Eq. (\ref{p3}) as

\begin{equation}
\Delta E= \frac{3}{2}\,K\, T_{\circ}\,N\; \frac{N_{\rm w}}{N} \;  \left(\sqrt{t^2\theta}+\sqrt{t^2/\theta}\right).
\label{p4}
\end{equation}

If we go on to define a parameter $q$ as $N_{\rm w}/N$ and note that 
$\theta = (1/q) - 1$ we obtain the form:

\begin{equation}
\Delta E= E_{\circ} \, \sqrt{t^2}\, \sqrt{\frac{q}{1-q}}
\label{p5}
\end{equation}

where $E_{\circ}$ is the thermal energy of the region if all the gas is at the mean temperature $T_{\circ}$.\\
It is interesting to look at some qualitative inferences from Eq.~(\ref{p5}).
If we let the mass fraction in the warm phase become small, 
i.e. if we let $q \rightarrow 0$, the energy required to yield a specific value
of $t^2$ also tends to zero. This does not, however, suggest an easy
mechanism for obtaining high values of $t^2$ since the temperature of 
the warm phase would in this case tend to infinity, as we can see
here from:
\begin{equation}
\Delta T= T_{\circ} \, \sqrt{t^2}\, \sqrt{\frac{1}{q\,(1-q)}}
\label{p6}
\end{equation}

where $\Delta T$ is the temperature difference between the cool and the warm phase. There is little need to show that this is not a realistic case, but it is clear that if the warm phase gets so hot, with a density tending to zero and 
temperature tending to infinity, the recombination rate for  line emission would also tend to zero so that only the cool phase could be observed in line emission and no temperature fluctuations would be observed.\\
The other limiting case for q, i.e. $q \rightarrow 1$ is also not
a realistic case because we know that the observed values for $T_{\circ}$ are
close to the values predicted for models where the only heating 
mechanism is photoionization, so that the additional energy input 
cannot be a major fraction of the total. There is one case that is
useful to present in this initial consideration of the model: that 
for which $q \sim 1/2$ , i.e. where half the material is in each phase.
This minimizes the temperature difference between the two phases 
required to yield a given value for $t^2$ (which is the hypothesis implicitly assumed by Peimbert when he pointed out the implications of $t^2$, and is an implicit assumption underlying the derivation of $t^2$ from observation), and from Eq.~(\ref{p6}) the 
additional heating, over and above that from photoionization, is 
given by:
	
\begin{equation}
\Delta E_{\frac{1}{2}}= E_{\circ} \; \sqrt{t^2}
\label{p7}
\end{equation}

To summarize this section, we have looked at three limiting cases:
\begin{itemize}
\item $q\rightarrow 1$. All the gas is heated by a mechanism other than photoionization. We have shown that this can be ruled out
\item $q\rightarrow 0$. A small fraction of the gas is heated to very high temperature. We have shown that this is very unlikely to account for the $t^2$ values observed, but will discuss it further below.
\item $q=\frac{1}{2}$. This case has be analyzed assuming essentially the same density for the warm and cool phases, and Eq.~(\ref{p7}). This condicion is too restrictive and we will relax it in section \S \ref{s1}.
\end{itemize}

\subsection{The densities of the warm and the cool media are different.}
\label{s1}

In section \ref{s0} we have implicitly assumed as an initial
condition that the densities of the warm and the cool media are
the same. Although this can occur in selected conditions, in general
this is improbable, so we now generalize to include density differences
between the two. If we let the densities be $n_{\rm w}$ and $n_{\rm c}$ for the warm and cool medium respectively, we need to multiply $\theta$, in Eqs.~(\ref{p1}) and (\ref{p2}) by $n_{\rm c}/n_{\rm w}$ and in this case the relation between $q$ and $\theta$ remains unchanged. We then obtain for the energy $\Delta E$ required to produce  a measured value for the fluctuation parameter $t^2$.

\begin{equation}
\Delta E= E_{\circ}\; \sqrt{t^2}\; \frac{(1-q)\, n_{\rm c} +n_{\rm w}\, q}{\sqrt{n_{\rm c}\,n_{\rm w} }}\; \sqrt{\frac{q}{1-q}}
\label{p8}
\end{equation}

In the limit of small fluctuations  $q = \frac{n_{\rm c}}{n_{\rm c}+n_{\rm w}}$ and as in the previous case we have $\Delta T_{\rm min}= 2 T_{\circ} \, \sqrt{t^2}$ which yields:
\begin{equation}
\Delta E_{n{\rm c}/(n{\rm c}+n{\rm w})}= 2\; E_{\circ}\; \sqrt{t^2}\; \frac{ n_{\rm c}}{n_{\rm c} +n_{\rm w}}
\label{p9}
\end{equation}

As $n_{\rm w}$ is always finite and non-zero we find  that in the general
case $\Delta E_{n{\rm c}/(n{\rm c}+n{\rm w})} < 2\; E_{\circ}\; \sqrt{t^2} $.\\

\subsubsection{The case of pressure equilibrium.}

It is worth setting out in detail the conditions implied by
pressure equilibrium between the two phases as many models of the ISM,
following Spitzer (1978), take this as the basis for determining
the parameter relations between phases. For pressure equilibrium
the relation between $n_{\rm w}$ and $n_{\rm c}$ is:

\begin{equation}
n_{\rm c} T_{\rm c}= n_{\rm w} (T_{\rm c}+\Delta T) \Rightarrow  n_{\rm w}= \frac{n_{\rm c}}{1+\frac{\Delta T}{T_{\rm c}}}
\label{p10}
\end{equation}

For small temperature fluctuations $q$ then becomes:

\begin{equation}
q =  \frac{T_{\rm c}+\Delta T}{2 T_{\rm c}+\Delta T}
\label{p11}
\end{equation}

 and this in turns leads to

\begin{equation}
\Delta E_{n{\rm c}/(n{\rm c}+n{\rm w})}= 2\; E_{\circ}\; \sqrt{t^2}\;\frac{T_{\rm c}+\Delta T}{2 T_{\rm c}+\Delta T}
\label{p12}
\end{equation}

and if we recall that $\Delta T = 2 T_{\circ} \sqrt{t^2}$ we clearly find that for small fluctuations of temperature, under pressure equilibrium, $ \Delta E_{n{\rm c}/(n{\rm c}+n{\rm w})} \sim E_{\circ}\sqrt{t^2}$.\\

\subsubsection{The case of large temperature difference}

The limiting case for large temperature differences occurs 
when $q \rightarrow  0$, and in that case the density of the warm phase also $\rightarrow 0$.
For that case, Eq.~(\ref{p8}) will give us:

\begin{equation}
\Delta E_{q \rightarrow 0}=  E_{\circ}\; \sqrt{t^2}\; \sqrt{ \frac{n_{\rm c}}{n_{\rm w}}\; q}
\label{p13}
\end{equation}

Bearing in mind that $q$ is the ratio between the warm gas mass and the 
total mass we can express this as $q= n_{\rm w} V_{\rm w}/(n_{\rm w}V_{\rm w}+n_{\rm c}V_{\rm c})$.\\
As we are always dealing with a finite volume, as $n_{\rm w} \rightarrow  0$ this becomes $q = \frac{n_{\rm w}V_{\rm w}}{n_{c}V_{\rm c}}$ so that Eq.~(\ref{p13}) can be written as:

\begin{equation}
\Delta E_{q \rightarrow 0}=  E_{\circ}\; \sqrt{t^2}\; \sqrt{\frac{V_{\rm w}}{V_{\rm c}}}
\label{p14}
\end{equation}
This is the excess energy which an ionized region requires in order to 
generate a given value of $t^2$ when the cool phase is much denser than
the warm phase.


\section{Application of the equations.}
\subsection{Considerations for  $q \rightarrow  0$: the hot phase.}

The situation described in Eq.~(\ref{p13}) corresponds to a 
model in which the two emitting phases have a strong contrast in 
density. This could, hypothetically, be the case for a model of 
the ISM such as that of Cox and Smith (1974) or McKee and Ostriker
(1977) in which the ISM is formed by cool dense clumps of gas with
warm envelopes surrounded by a much hotter tenuous interclump 
substrate. These specific models invoke the presence of supernovae
within the ionizing star clusters to originate and maintain the hot
substrate, so that they apply principally to those highly luminous {\hii} regions which are ionized by a large OB star cluster. The characteristic densities and temperatures of this 
hot phase are of order  $n\sim 10^{-2.5}$~cm$^{-3}$ and $T \sim 10^{5.7}$~K (McKee \& Ostriker 1977) respectively. We can show using our model that these
values are unlikely to be consistent with the role of this hot phase
as the source of the observed temperature fluctuations. We can estimate
the energy, $\Delta E_{\rm q \rightarrow 0}$, required to cause the observed range of $t^2$ values, as the density tends to zero, which is a characteristic
property of the hot phase. The estimate is based on a ratio of the
volumes for the hot and cool phases, which can be found by estimating
the the geometrical filling factor $\phi_{\rm G}$ of the
cool phase, which can be determined spectroscopically with the
aid of a relevant theoretical model. We will use the estimate of
Giammanco et al. (2005) who give $10^{-3} \leq \phi_{\rm G} \leq 10^{-2}$.  
The ratio of the volumes of the hot and cool phases is just the
reciprocal of $\phi_{\rm G}$. Combining this range of values with a
characteristic value for $t^2$ of $0.01$ and substituting in Eq.~(\ref{p14})
we obtain $\Delta E_{\rm q \rightarrow 0} \geq E_{\circ}$. This is not a reasonable physical solution, because this would imply a source of additional energy more powerful than  photoionization, and this
would undoubtedly cause rapid heating of the cool component. However to show even more clearly that this is not a realistically observable case we take some further considerations into account. Using the relation $\Delta E_{\rm q \rightarrow 0} \geq E_{\circ}$, our definition of $E_{\circ}$, and knowing that the density of the cool phase is much greater than that of the hot  phase, we find:
	
\begin{equation}
\frac{3}{2} T_{\rm h} n_{\rm h} V>\frac{3}{2} T_{\circ} n_{\rm c}\; \phi_{\rm G} V
\label{p20}
\end{equation}

where we have relabelled our variables $T_{\rm w}$, $n_{\rm w}$, calling them 
$T_{\rm h}$ and  $n_{\rm h}$ in this case since we are dealing with a truly hot
rather than a merely warm higher temperature phase. Substituting the
appropriate values in (\ref{p20}) we find  $T_{\rm h}>\frac{n_{\rm c}}{n_{\rm h}} \phi_{\rm G} \geq 10^{5.5}$. The lower limit, which in fact does coincide with the temperature for the hot phase proposed by McKee and Ostriker (1977) is obtained using a value for $\phi_{\rm G}$ of 10$^{-3}$. Under this condition, however, the excess energy $\Delta E$ will be three times greater than the mean
energy $E_{\circ}$, as can be inferred from Eq.~(\ref{p14}), and this is not
physically realistic. In fact, as shown in Giammanco et al. (2004)
values of $\sim 10^{-3}$ for $\phi_{\rm G}$ are considerably less than the true filling factor of the cool medium and represent, rather, the part of the
cool medium which is sufficiently photoionized to emit significant
line radiation. On the other hand for the excess energy to
be less than $E_{\circ}$ the filling factor $\phi_{\rm G}$  would have to rise to values $\gtrsim 10^{-2}$, a more realistic value according to Giammanco et al. (2004), but in that case the temperature of the hot phase would
have to be almost an order of magnitude higher than that predicted
by the McKee--Ostriker model, which is again not plausible. So we
are forced to conclude that the hot phase in this type of models cannot
be the source of the observed temperature fluctuations, even for
highly luminous {\hii} regions with cool clumps embedded in a much
hotter substrate. For these objects we could turn to the mechanisms
proposed in Giammanco \& Beckman (2005): cosmic ray ionization of
the cool material, or consider that the same mechanism is acting
as acts in smaller less luminous regions, such as the Orion nebula,
which therefore could be magnetic field reconnection as postulated
below.

\subsection{The excess energy to sustain the $t^2$ observed in the Orion Nebula}

Having derived expressions for the excess energy of the warm phase we
can examine their applicability by substituting into them characteristic
observationally derived values applying reasonable physical constraints.
Although at first sight the case where the two phases have the same
density may not appear to be physically plausible we will show here
that it is in fact a fair paradigm. Note that we are considering a 
medium which is virtually fully photoionized, whose temperature is 
determined to first order by the equilibrium between heating due to 
photoionization and cooling by the emission of forbidden line radiation.
In addition we must assume that some extra heating mechanism can 
inject an additional flux of energy which heats the gas incrementally
to second order. The above calculations do not give us information 
about the spatial distribution of the cool and warm media, which we must
try to infer from observations. From a map of the temperature 
distribution obtained by O'Dell et al. (2003) we know that for 
the nearby well resolved Orion nebula the volume elements, each having
a uniform temperature, are distributed uniformly within the nebula, and 
have size scales smaller than 1 arcsec, which corresponds to 
$\sim 2.5 \times 10^{-3}$~pc using a distance to the nebula of 500~pc. The fact that there is no obvious distinction between the sizes of the warm and cool
elements implies that their interfaces are distributed equally 
within the nebula. Transforming this to a condition on the volumes
of the two phases suggests that the value of $q$ must be close to $1/2$,
which as we have shown above is a good situation for the development
of temperature fluctuations.\\
In order to quantify the relevant effects it is useful to work 
with the energy density: $e_{\circ}$ the energy per unit volume within a region.
This is expressed for the mean values of the parameters concerned as
$e_{\circ}=E_{\circ}/V=\frac{3}{2}\; n_{\rm e}\; K T_{\circ}$. \\
For the Orion nebula, using observationally derived values of $n_{\rm e} = 10^4$~ cm$^{-3}$ (Ferland, 2001, Estebam et al., 2004) and $T_{\circ} = 10^4$~K 
(Ferland, 2001) we find  $e_{\circ}\sim 2\times 10^{-8}$~erg~cm$^{-3}$. In order to compute the energy excess in elements of the warm phase, noting that
these occupy one half of the volume, we use Eq. (\ref{p7}) which gives $\Delta e= e_{\circ} \sqrt{t^2}$. For a value for $t^2=0.022$ determined
observationally for Orion by Esteban et al. (2004) we derive  $\Delta e\sim 3\times 10^{-9}$  erg/cm$^{-3}$\\
For two phases in pressure equilibrium the estimate will change
very little, since for small fluctuations Eq.~(\ref{p10}) gives $n_{\rm w} \sim n_{\rm c}$.
We will use this condition to calculate the energy required of any 
physical mechanism which heats the cool medium, converting it to 
the warm phase. For pressure equilibrium we need to include the 
component of energy required to expand the gas at constant pressure,
which is $P\;\Delta V$, where $\Delta V$ is the corresponding volume 
increment. From the basic assumption of particle conservation we have 
		
\begin{equation}
\Delta V =V_{\circ}^{\rm w}\; \left( \frac{n_{\rm c}}{n_{\rm w}}-1 \right)
\label{p15}
\end{equation}

where $V_{\circ}^{\rm w}$ is th volume occupied by the warm component before it
expands, when its density is  $n_{\rm c}$. Using the $q$ parameter we can
write:

\begin{equation}
q=\frac{N_{\rm w}}{N}=\frac{n_{\rm c}V^{\rm w}_{\circ}}{n_{\rm c}V^{\rm w}_{\circ}+n_{\rm c}(V-V^{\rm w}_{\circ})}
\label{p16}
\end{equation}

so that we derive:

\begin{equation}
V^{\rm w}_{\circ}= q\; V
\label{p17}
\end{equation}

Substituting (\ref{p17}) in (\ref{p15}) and including expression (\ref{p10}) we have

\begin{equation}
\Delta V = q\; V \; \frac{\Delta T}{T_{\rm c}}.
\label{p18}
\end{equation}
Using the standard form of the ideal gas equation we can write $P_{\rm e}= K\, n_{\rm c}\, T_{\rm c}$, and as we know that for pressure equilibrium, and for small fluctuations $q \sim 1/2$, $n_{\rm c} \sim n_{\rm n}$ and $\Delta T= 2 T_{\circ} \sqrt{t^2}$ we find

\begin{equation}
P \Delta V \sim \frac{2}{3}\; E_{\circ}\; \sqrt{t^2}.
\label{p19}
\end{equation}

We need to add this term to the excess thermal energy required to 
produce the warm phase. So for the value of $t^2$ observed in the Orion nebula, assuming pressure equilibrium and small temperature fluctuations we find that the total energy density needed to heat the gas to $T_{\rm w}$ will be
  $\sim e_{\circ} \sqrt{t^2}+\frac{2}{3} e_{\circ} \sqrt{t^2}$ which corresponds to $5 \times 10^{-9}$~erg~cm$^{-3}$.

\subsubsection{The energy balance: temporal considerations.}
Up to this point we have considered the excess energy which 
needs to be injected into the warm or hot phase in order to yield
the observed range of values for $t^2$. It is, however, well known that
even at the low densities which prevail in the regions of the ISM under
consideration collisionally excited emission will produce relatively 
rapid cooling so that the warm phase will quickly relax to the 
mean temperature.  An estimate of the cooling function was given by 
Ferland (2001) for the region of the Orion nebula in which {\oiii} is
present. Using this function the excess energy required to yield the 
conditions observed in the Orion nebula ($\Delta e\sim 3 \times 10^{-9}$ erg/cm$^{-3}$, $\Delta T= 2 T_{\circ} \sqrt{t^2}$), we find an estimate for the typical cooling time  $\tau \sim 92 \times (10000/n_e)$  days (this estimate was obtained using the cooling and heating functions presented in fig 5. of Ferland 2001, and Osterbrock, 1989). If there  were no continued energy supply a warm element of the ISM would dissipate its temperature excess in times of this order. We can consider this  timescale in the framework of two alternative scenarios: either an energy source which supplies a steady energy exces to a fixed set of volume elements, or a mechanism which can heat different sets of elements randomly.
Where approximately a half of the gas mass has the excess temperature
required to produce a given value of $t^2$, the injection timescale would
be around twice the cooling timescale.\\
It should be possible to distinguish between the two proposed
scenarios by repeatedly making parameter maps of an ionized nebula
with spatial resolution sufficient to tell whether or not the map shows
invariant zones of higher and lower temperature, or whether in a region where the mean value of $t^2$ does not vary with time, the local zones of 
raised temperature vary from observation to observation. Observations 
spaced in time at intervals of several months would be useful in 
distinguishing between the two scenarios for the Orion nebula, following
the technique of mapping the emitting object in a set of selected 
emission lines, as carried out by O'Dell et al. (2003).

\subsubsection{A specific mechanism: magnetic field reconnection.}

We showed in section \S 2.1 that for the case of equal densities 
 of the warm and cool phases, for the value of $t^2$ measured by 
 Esteban et al. (2004) in the Orion nebula an excess energy of 
 $3 \times 10^{-9}$~erg~cm$^{-3}$ would be required to maintain the observed 
 temperature difference between the two phases. One plausible source
 for this excess energy is that supplied by magnetic fields, and in 
 particular by the local reconnection of turbulent fields which can
 be continually renewed within this turbulent ionized medium. 
 Quantitatively the excess energy density required in the Orion nebula
 as specified here could be supplied by the destruction of a $B$ field
 of amplitude $\sim 300~\mu$G, which is not an unreasonable magnitude taken
 in comparison with the rather few relevant measurements available.
 For the case of pressure equilibrium between the two phases
described in section \S 3.2 and taking into account the full expenditure
 of energy in expanding the warmer phase as well as in heating it
 a rather larger field is required, since the excess energy in the 
 warm phase is somewhat greater. However this extra energy corresponds
 to a field of order $\sim 350~\mu$G, again not an exorbitant value. A more
 realistic way of approaching this scenario is to consider that a 
 fraction of the mean magnetic field in the gas is destroyed per unit
 time by connection. In a recent article Beckman \& Rela\~no (2004)
 estimated that if there is equipartition of energy between the kinetic
 and magnetic energies in a typical {\hii} region, the required turbulent
 field strengths would be of order a few tens to a few hundred gauss.
 The equipartition hypothesis was considered also by Ferland (2001), who made the suggestion that magnetic field could give rise to temperature fluctuations for the Orion nebula. Based on the measured broadening of the emission lines from the Orion nebula (Casta\~neda, 1998, O'Dell, 2000) he estimated a magnetic field of $\sim 400~\mu$G. Also of some relevance in this contest is the detection by Abel et al. (2004) of a turbulent field of amplitude $100~\mu$G in the neutral hydrogen veil between the Orion nebula and ourselves.
 Although in a different regime, this is a useful pointer to the 
 order of magnitude of the IS fields in the nebula. The observation of O'Dell et al. (2002) suggest that an element of the nebula at a single coherent temperature  has a characteristic scale of the turbulent elements within the nebula. Casta\~neda et al. (1998) found that, at the limit of their ground based resolution turbulent structure could be identified on scales of around 1 arcsec, which is in fair agreement with hypothesis of turbulent magnetism as the origin of the temperature fluctuations. 
 
 \section{Conclusions}
 
We have examined the energy balance in {\hii} regions with the 
specific focus of constraining models which can account quantitatively for 
their observed temperature fluctuations. As we addressed in a previous 
article the situation in highly luminous regions, where dense cool clumps 
are embedded in a warm medium, we have concentrated here on smaller 
more homogeneous situations, typified by the ionized gas in the Orion 
Nebula for which detailed observations of the temperature inhomogeneity 
are available. Based on a simple formalism for a two component structure, 
which deals with the energy required to produce components with a significant temperature difference between them, we find that the most plausible general scenario for this case is of two components of similar density and occupying comparable fractions of the total volume. A possible mechanism for producing this structure is examined using the observations of the temperature fluctuations in the Orion Nebula by O'Dell et al.~(2003). We find that reconnections in a magnetic field induced by the turbulence within the ionized gas are capable of supplying sufficient energy if the mean induced field strength is of order a few hundred $\mu$G. If these reconnections occur at random within the ionized volume the fluctuations should have a decay timescale similar to the radiative dissipation rate exhibited by gas whose temperature has been raised to a differential value from the mean corresponding to that measured by the temperature fluctuation parameter. This difference is of order 2000~K in a photoionized medium of mean temperature 10000~K, and the corresponding dissipation timescale is computed as being a few months for conditions in the Orion Nebula. This prediction is in principle testable observationally.

\begin{acknowledgements}
This work  was carried out with the help of grants  AYA 2004-08251-CO2-01 of the Spanish Ministry of Education and Science, and P3/86 of the Instituto de Astrof\'{\i}sica de Canarias. We are gratefull to Manuel Peimbert and to Valentina Luridiana for valuable discussions.
\end{acknowledgements}

\end{document}